\documentclass[journal abbreviation]{copernicus2}

\begin{document}

\title{Generalised partition functions: Inferences on phase space distributions%by seed-field generation
}

\author[1]{R. A. Treumann\thanks{Visiting the International Space Science Institute, Bern, Switzerland}}
%\author[3]{R. Nakamura}
\author[2]{W. Baumjohann}
%\author[2]{Y. Narita}

\affil[1]{Department of Geophysics and Environmental Sciences, Munich University, Munich, Germany}
%\affil[2]{Department of Physics and Astronomy, Dartmouth College, Hanover NH 03755, USA}
\affil[2]{Space Research Institute, Austrian Academy of Sciences, Graz, Austria}
%% The [] brackets identify the author to the corresponding affiliation, 1, 2, 3, etc. should be inserted.

\runningtitle{Partition-Function}

\runningauthor{R. A. Treumann and W. Baumjohann 
}

\correspondence{R. A.Treumann\\ (rudolf.treumann@geophysik.uni-muenchen.de)}

\received{ }
%\pubdiscuss{ } %% only important for two-stage journals
\revised{ }
\accepted{ }
\published{ }

%% These dates will be inserted by the Publication Production Office during the typesetting process.

\firstpage{1}

\maketitle

%\begin{abstract}
\subsection*{\bf{Abstract}} It is demonstrated that the statistical mechanical partition function can be used to construct various different forms of phase space distributions. This indicates that its structure is not restricted to the Gibbs-Boltzmann factor prescription which is based on counting statistics. With the widely used replacement of the Boltzmann factor by a generalised Lorentzian [also known as the $q$-deformed exponential function, where $\kappa=1/|q-1|$, with $\kappa, q \in\textsf{R}$] both the Kappa-Bose and Kappa-Fermi partition functions are obtained in quite a straightforward way, from which the conventional Bose and Fermi distributions follow for $\kappa\to\infty$. For $\kappa\neq\infty$ these are subject to the  restrictions that they can be used only at temperatures far from zero. They thus, as shown earlier, have little value for quantum physics. This is reasonable, because physical $\kappa$-systems imply strong correlations which are absent at zero temperature where appart from stochastics all dynamical interactions are frozen. In the classical large temperature limit one obtains physically reasonable $\kappa$-distributions which depend on energy respectively momentum as well as on chemical potential. Looking for other functional dependencies, we examine Bessel functions whether they can be used for obtaining valid distributions. Again and for the same reason, no Fermi and Bose distributions exist in the low temperature limit. However, a classical Bessel-Boltzmann distribution can be constructed which is a Bessel-modified Lorentzian distribution. Whether it makes any physical sense remains an open question. This is not investigated here. The choice of Bessel functions is motivated solely by their convergence properties and not by reference to any physical demands. This result suggests that the Gibbs-Boltzmann partition function is fundamental not only to Gibbs-Boltzmann but also to a large class of generalised Lorentzian distributions as well as to the corresponding nonextensive statistical mechanics. 
%\end{abstract}

\section{Introduction}
Since its introduction by \citet{vasyliunas1968}\footnote{\citet{vasyliunas1968} acknowledges that in applying the Kappa distribution  as an apparently useful fit to the observed energy dependence of low-energy electron fluxes in the geomagnetic-tail plasma sheet he followed a suggestion of its functional form by Stanislaw Olbert.}, the so-called Kappa-distribution function $f_\kappa(x)\propto[1+x/\kappa]^{-(\kappa+r)}$ has experienced increasing attention and application in space plasma problems.\footnote{For a recent compilation and in-depth discussion of the various aspects and applications of the Kappa distribution the reader is referred to the extended presentations contained in \citet{livadiotis2009,livadiotis2013} as well as to the almost complete list of papers referenced therein. This list gives a historical record of the work done on and application of the Kappa distribution as well as its relation to the celebrated Tsallis nonextensive thermostatistics \citep{tsallis1988,tsallis1998,gellmann2004}.} The Kappa distribution turned out to fit not only the geotail low-energy electron distribution sufficiently well but also the fluxes of energetic ions in the tail \citep[cf., e.g.,][]{christon1988} which demonstrates that the Kappa distribution applies successfully to physical problems even though its  physical origin was not entirely clarified.  It has also been used in several formal contexts including $q$- or $\kappa$-generalisations \citep[cf., e.g.,][]{lenzi1999,treumann2014a,treumann2015} of various mathematical functions and functional transforms (see also the reference lists of papers cited in footnote 2). As for an example, even modified Feynman path integrals have been defined based on generalised Lorentzians \citep{treumann1998}.

In principle, the Kappa distribution is a probability distribution function which, mathematically, is identical to the generalised Lorentzian respectively $q$-deformed exponential \citep[cf., e.g.,][]{livadiotis2009}. The parameter $q\in\textsf{R}$ had been introduced first by \citet{renyi1955,renyi1970} as power of a $q$-generalised logarithmic Boltzmann entropy that found wide application in the theory of deterministic chaos and the related thermodynamics \citep[cf., e.g.,][]{beck1997}. \citet{tsallis1988} referred to it in postulating his non-extensive, conveniently simple version of entropy which became the basis of the celebrated Tsallis-nonextensive thermostatistics \citep[cf. also][for some rectifications of his earlier work]{tsallis1998}.

Formal relations between the Tsallis statistical mechanics of non-extensive entropies and the Kappa-distribution do indeed exist. This is not surprising,  because the parameter $\kappa=1/|q-1|$ can almost trivially be related to the parameter $q$ that appears in the non-extensive thermostatistics. This relation was first implicitly used in a note investigating superdiffusion near the magnetopause \citep[][see the appendix of that note]{treumann1997} when referring to L\'evy-flight statistics in the form proposed by \citet{shlesinger1987}\footnote{Its most recent exposition is found in \citet{zaburdaev2015}.} though not referring to Tsallis' non-extensive statistics such that the coincidence was somehow accidental. It was independently elaborated by \citet{milovanov2000}, \citet{leubner2002} and others in various contexts.\footnote{For the complete lists of references see again \citet{livadiotis2009,livadiotis2013} and \citet{livadiotis2015}.} It should, however, be noted that the correct relation of the Kappa distribution to the physical temperature in the plasma was first given  on thermodynamic reasons in \citet{livadiotis2009} and also confirmed from a rigorous calculation of the particular case of the time-asymptotic (stationary) electron distribution resulting in the interaction of electrons with Langmuir waves \citep{yoon2012}.

A heuristic generalisation of statistical mechanics to general entropies has been proposed more recently \citep{treumann2014b} based on the fundamental Gibbs prescription of relating any entropy to the differential phase-space element d$\Gamma$. With the entropy being a functional of the energy that theory states that it is possible to derive a general expression for the probability of occupation of physical states. This requires knowledge of the {inverse entropy functional} which in most cases will be difficult to construct. In the particular case of the generalised Lorentzian it was shown there that the inverse functional can indeed be obtained. It turns out that in this case it is identical to what in Tsallis' nonextensive statistical mechanics is called ``escort distribution" \citep{beck1997,gellmann2004}. It not only led to the reproduction of the Kappa-distribution as a physically accessible distribution function but also to make it consistent with statistical mechanics. This generalisation was made possible because of the familiar additional prescription used in the definition of the generalised Lorentzian (or $\kappa$-modified exponential) that in the limit $\kappa\to\infty$ the statistical mechanics should reproduce Gibbs' statistical mechanics. This is a severe additional constraint that might not be satisfied nor necessary in  any other choice of the functional which replaces the exponential or the generalised Lorentzian. Any physical constraints are not expected to merge the Gibbs-Boltzmann case except in the absence of all correlations and complete stochasticity. Rather  they are the requirement of reproducing the thermodynamic relations in the stationary state \citep[as done by][for the Kappa distribution]{livadiotis2009} -- if only it exists. 

It is interesting that the Kappa-distribution understood as a probability distribution also reproduces distributions that are obtained when analysing intermittency\footnote{As for a typical example of intermittence in solar wind magnetic turbulence see, for instance,  \citet{brown2015}.} in the data of chaotic processes. In these cases it  sometimes properly maps the tails of the probability distributions allowing  for the determination of the power index $\kappa$. The physical reason for its occurrence can indeed be found in the deterministic chaos underlying the occurrence of intermittency. That, nevertheless, it can be related to Gibbs' statistics lets one ask whether one could go one step deeper in its foundations. It is known that no counting statistics exists which could reproduce the generalised Lorentzian statistical mechanics. What, however, if we ask for the Gibbsian partition function? To what extent does the Gibbsian partition function reproduce $\kappa$-distributions as distributions of physical states?

In the following we start from the general Gibbs-Boltzmann partition function as the accepted physical basis of statistical mechanics. We then transform it into a Kappa partition function and proceed to the derivation of the equation of state and the physical distribution of occupation of states. In doing so we follow the prescription of statistical mechanics in deriving the physical distribution function. The idea is thus very simple. However, this process is physically motivated and provides some additional physical insight. 

\section{Formulation}
The grand partition function $\mathcal{Z}_G$ that results from Gibbsian counting statistics can be written in the canonical form
\begin{equation}
\mathcal{Z}_G(\mu,V,T)=\prod_\mathbf{p}\Big\{\sum_n\big[\exp\beta(\mu-\epsilon_\mathbf{p})\big]^n\Big\}
\end{equation}
where $n=0,1,2,\dots$ is the occupation number of states \citep[cf., e.g.,][]{kittel1980,huang1987}, and $\epsilon_\mathbf{p}=\epsilon(\mathbf{p}),\: \mu$ are the respective energy as function of momentum $\mathbf{p}$ and chemical potential, the latter being a function of density. $\beta$ is the inverse kinetic temperature, with the latter taken in energy units. The summation refers to all  $n$. Clearly the sum of all occupations is the total particle number $\mathcal{N}$. It is stressed that this expression holds under the assumption that the basic process that underlies its derivation is purely stochastical. It is based on throwing coins and counting the statistical outcome of how they become distributed over the available number of boxes in phase space. Any correlations are excluded. Respecting a different statistics like, for instance, Bayesian statistics which satisfies certain conditions, is excluded. This stochasticity is responsible for the presence of the exponential function, i.e. a Gaussian probability distribution..

We now violate, on this advanced level, the stochastic assumption. We assume that the structure of the partition function will remain intact if we replace the exponential with another function that in some limit reproduces the exponential. Such a function is, as for an example, the $\kappa$-generalised Lorentzian which has been used in several of the above cited publications (and referenes therein).

The above version of the partition function can also be written in another form
\begin{equation}
\mathcal{Z}'_G(\mu,V,T)=\prod_\mathbf{p}\Big\{\sum_n\exp \big[n\beta(\mu-\epsilon_\mathbf{p})\big]\Big\}
\end{equation}
which we indicate by a prime. On the Gibbsian level the two versions are identical because raising an exponential to power $n$ is the same as multiplying its argument by $n$.

Now we violate the assumption of pure stochasticity by introducing the Lorentzian replacing $\exp(ax)\to(1-ax/\kappa)^{-(\kappa+r)}$, where $0<\kappa\in\textsf{R}$ is some free parameter, and $0<r$ is a fixed number that has to be adjusted to satisfying the thermodynamic relations. Determination of $r$, for the classical case, has been done in several places \citep[e.g.,][]{yoon2012,livadiotis2013,treumann2014b}.  Then we obtain two new versions of the partition function
\begin{equation}
\mathcal{Z}_\kappa(\mu,V,T)=\prod_\mathbf{p}\Big\{\sum_n\big[1-\beta(\mu-\epsilon_\mathbf{p})/\kappa\big]^{-n(\kappa+r)}\Big\}
\end{equation} 
and
\begin{equation}
\mathcal{Z}'_\kappa(\mu,V,T)=\prod_\mathbf{p}\Big\{\sum_n\big[1-n\beta(\mu-\epsilon_\mathbf{p})/\kappa\big]^{-(\kappa+r)}\Big\}
\end{equation}
There is a big difference between these two versions in the position of the occupation number $n$. In the second form it is located inside the argument of the Lorentzian. This inhibits any further analytical treatment by summing the partition function up except in the case of a Fermi system which we therefore treat first. 

\section{Fermi partition function analysis}
In a Fermi system we can only have two occupations $n=0$ and $n=1$. With this restriction we find for either of the above versions \begin{equation}
\mathcal{Z}^F_\kappa(\mu,V,T)=\prod_\mathbf{p}\Big\{1+\big[1-\beta(\mu-\epsilon_\mathbf{p})/\kappa\big]^{-(\kappa+r)}\Big\}
\end{equation} 
Incidentally identically the same result is obtained from the second form of the partition function in this case. Thus there is no difference in a Fermi system between the effect of the correlations introduced by changing from the Gibbs exponential to the generalised Lorentzian.

From the partition function one obtains the ideal gas equation of state as 
\begin{equation}
\frac{PV}{T}=\log\mathcal{Z}^F_\kappa=\sum_\mathbf{p}\log\Big\{1+\big[1-\beta(\mu-\epsilon_\mathbf{p})/\kappa\big]^{-(\kappa+r)}\Big\}
\end{equation}
More interesting is the average occupation number $\langle n_\mathbf{p}\rangle^F_\kappa$ of states which is prescribed by the partition function. It follows from the negative derivative of the logarithm of the partition function
\begin{equation}
\langle n_\mathbf{p}\rangle^F_\kappa=-\frac{1}{\beta}\frac{\partial}{\partial\epsilon_\mathbf{p}}\log\mathcal{Z}_\kappa^F
\end{equation}
A simple calculation then yields that the Fermi-Kappa-distribution becomes
\begin{equation}
\langle n_\mathbf{p}\rangle^F_\kappa=\frac{1+r/\kappa}{[1-\beta(\mu-\epsilon_\mathbf{p})/\kappa]}\left\{1+\left[1-\beta(\mu-\epsilon_\mathbf{p})/\kappa\right]^{\kappa+r}\right\}^{\!\!\!-1}
\end{equation}
This is the distribution we have obtained earlier in \citet{treumann2014b} and already before. Notably it is not the distribution which one would obtain by simply replacing the exponential function in the common Fermi distribution by the corresponding generalised Lorentzian! 

For $\kappa\to\infty$ the last expression becomes the ordinary Fermi distribution. This can be easily checked. However, for finite $\kappa<\infty$ it has no zero temperature limit. At $T=0$ no states can be occupied. This is very satisfactory because at zero temperature there is no mechanism that could generate any correlations. Hence, the above Fermi-Kappa distribution has meaning only at finite temperature, and one has that the chemical potential $\mu<0$ in all cases. On the other hand, at fixed $\kappa$ and high temperature one simply recovers the ordinary Kappa distribution. We may note that here we are dealing with the ideal gas. In non-ideal gases one would add the external or internal potential fields to the energy which causes a shift in the energy scale and would lead to additional effects which are not included here. One may note that external potentials and therefore energy shifts may cause observable effects. 

\section{Bosonic distribution}
For the Boson distribution we refer to the function $\mathcal{Z}_\kappa$ which can be summed up over $n$. The result is trivially given by
\begin{equation}
\mathcal{Z}^B_\kappa(\mu,V,T)=\prod_\mathbf{p}\Big\{1-\big[1-\beta(\mu-\epsilon_\mathbf{p})/\kappa\big]^{-(\kappa+r)}\Big\}^{-1}
\end{equation} 
Accordingly the bosonic ideal gas equation of state is found as
\begin{equation}
\frac{PV}{T}=\log\mathcal{Z}^B_\kappa=-\sum_\mathbf{p}\log\Big\{1-\big[1-\beta(\mu-\epsilon_\mathbf{p})/\kappa\big]^{-(\kappa+r)}\Big\}
\end{equation}
and the average bosonic occupation number of Bose distribution becomes
\begin{equation}
\langle n_\mathbf{p}\rangle^B_\kappa=\frac{1+r/\kappa}{1-\beta(\mu-\epsilon_\mathbf{p})/\kappa}\left\{1-\left[1-\beta(\mu-\epsilon_\mathbf{p})/\kappa\right]^{\kappa+r}\right\}^{\!\!\!-1}
\end{equation}
The symmetries between the Bose-Kappa and Fermi-Kappa cases are striking. Again, the Bose-Kappa distribution has no zero temperature limit. Like the Fermi-Kappa distribution it exists only at sufficiently high or simply finite temperatures. Its high energy limit is the ordinary Kappa distribution with negative chemical potential $\mu<0$ -- one may note that for large $\epsilon_\mathbf{p}$ the negative signs in the denominator cancel.

\section{Classical limit}
For high-temperature high-energy classical gases both distributions above become a reasonable classical limit known as the Kappa Distribution. In such classical cases the chemical potential becomes negative. Then the complete classical distribution that is in accord with the partition function assumes the form
\begin{equation}
\langle n_\mathbf{p}\rangle_\kappa=\left(1+\frac{r}{\kappa}\right)\Big[1+\beta\big(|\mu|+\epsilon_\mathbf{p}\big)/\kappa\Big]^{-(\kappa+r+1)}
\end{equation}
This occupation number is still subject to normalisation to the total particle density $N=\mathcal{N}/V$ and adjustment of the index $r$ to thermodynamics. We noted that this has been done in different ways \citep{livadiotis2009,livadiotis2013,yoon2012,treumann2014b} yielding $r+1=\frac{5}{2}$. Normalisation requires integration over the phase space volume. 

We note in passing that the relativistic equivalent of the above Kappa distribution should become
\begin{equation}
\big\langle n_{\gamma} \big\rangle_\kappa= \left(1+\frac{\rho}{\kappa}\right)\Big\{1+\beta_r\Big[|\mu_r|+\gamma(\mathbf{p})\Big]/\kappa\Big\}^{-(\kappa+\rho+1)}
\end{equation}
where $\gamma(\mathbf{p})=\sqrt{1+p^2/m^2c^2}$ is the relativistic energy factor, and $\beta_r=mc^2/T$,  $\mu_r=\mu/mc^2$ are the normalised inverse temperature respectively chemical potential. The relativistic exponent $\rho$ differs from its non-relativistic counterpart $r$. It must be adjusted by satisfying the relativistic thermodynamic relations \citep[cf., e.g.,][]{treumann2014b}.

It is interesting that the chemical potential cannot be extracted from this expression. This makes its use as a physical distribution more difficult and requires use of approximation methods to eliminate $\mu$. This must be done by standard procedures referring to the density as a known quantity \citep[cf., e.g.,][]{huang1987}. In space-plasma applications the Kappa-distribution is used as a probability, and it is assumed that $\mu=0$ which implies that the particles under consideration behave like massless Bosons. 

The straightforward calculations by \citet{yoon2012} in highly diluted high temperature plasmas seem to confirm this assumption at least in the interaction of electrons with Langmuir waves, the case investigated there. \citet{yoon2012} included spontaneous and induced emissions, scattering and absorption of Langmuir waves when determining the shape of the electron distribution function in final stationary equilibrium. These processes seem not to generate any chemical potential at given particle number respectively density. 

A negative chemical potential which is expected in the classical case should cause trapping and thus retarding the electrons respectively accumulating them around the trapping potential, i.e. the chemical potential. This is obviously not the case! Scattering of electrons by absorbing wave momentum and energy pushes the electrons instead  into the extended tail of the Kappa distribution. It thus overcompensates for the chemical potential that might have been produced by the retardation effect related to the spontaneous and induced  emissions. Hence, in this particular case one encounters that statistical mechanics acts self-compensating for the chemical potential while generating the power law tail on the distribution. Since entropy is increased hereby, the process of generation of the tail seems favorable for the interaction. One might conclude that the Kappa distribution and its related statistical mechanics strictly apply to conditions only when the chemical potential is suppressed. Such conditions seem, however, to be realised quite frequently.

We note that this effect had already been observed earlier in a model where electrons were put into a heat bath of radiation photons \citep{hasegawa1985}. Clearly the photon distribution has zero chemical potential. Similarly, the Langmuir photon distribution has zero chemical potential. 

These observations as well as the results of the rigorous calculations are important. They suggest that in any Fermi-like process leading to the formation of energetic tails on the particle distribution, the chemical potential will be vanishingly small.  

This observation also explains why the particle spectra measured by \citet{vasyliunas1968} and \citet{christon1988} all obeyed almost perfect Kappa-distributions. And any cosmic ray spectra that extend over many orders of magnitude are probably simple power laws for the same reason: they result from scattering while themselves contributing to the photon spectrum by spontaneous emission  and attribution of a tiny fraction of energy in photons only that is insufficient to produce a sufficiently strong negative chemical potential that could suppress their runaway into the energetic tail. Tail generation is obviously entropically favoured over both heating and radiation.

\section{Preliminary discussion}
There is no known counting statistics in cases where the system is not stochastic but respects some internal correlations. It is not clear how such cases should be treated even then when the correlations have been specified from the very beginning. Possibly application of Bayesian statistics could offer a route to such systems. Statistical mechanics, however, seems not to have had any needs so far in non-stochastic  states on the microscopic level. These are usually treated by numerical simulations or kinetic theory where the evolution of the one-particle distribution function is followed in time. This is clearly the right physical approach to non-stationary systems in evolution. Statistical mechanics just deals with the stationary state of a system. 

That the introduction of correlations via the replacement of the exponential by the Lorentzian on the level of the partition function nevertheless reproduces the correct $\kappa$-statistics as derived intuitively from assumptions that have nothing in common with stochasticity, suggests that the structure of the partition function is more general than purely stochastic. It is just the sum over all occupations in the probabilities of states -- quite a general notion. One may thus ask whether not other functions exist with the physical meaning that they include correlations when used in the partition function. The requirement on them implies that they should behave correctly at large energies, i.e. converge for $\epsilon_\mathbf{p}\to\infty$. Moreover, in this limit they should possibly turn over to become gaussians. In the following we try such a case.

\section{Gibbsian-Bessel partition functions} 
A particular function which seems to offer itself is the modified Bessel function of the first kind $I_\nu(z)=\mathrm{e}^{-i\nu\pi/2}J_\nu(iz)$. An integral representation of this function is
\begin{equation}
I_\nu(z)=\frac{(z/2)^\nu}{\sqrt{\pi}\Gamma(\nu+\frac{1}{2})}\int_0^\pi \mathrm{d}\theta\ \mathrm{e}^{\pm z\cos\theta}\sin^{2\nu}\theta,\quad \nu>-\frac{1}{2}
\end{equation}
It converges for $z\to0$. Its asymptotic expansion for $z\to\infty$ is $I_\nu(z)\sim\mathrm{e}^{z}/\sqrt{z}$ and diverges for positive $z$.  For negative argument $-\frac{1}{2}\pi<\arg z< \frac{3}{2}\pi$ we have $I_\nu(-z)=\mathrm{e}^{i\nu\pi}I_\nu(z)$, which converges but may become complex depending on index $\nu$. Thus there are domains where it satisfies the primary need on a reasonable function that could possibly replace the Gibbs-Boltzmann exponential factor with the Gibbs-Boltzmann-Bessel factor. With this in mind we write
\begin{equation}
\mathcal{Z}_{GBB}(\mu,V,T)=\prod_\mathbf{p}\Big\{\sum_n\Big[I_\nu(z)\Big]^n\Big\}
\end{equation}
where we define $z=\beta(\mu-\epsilon_\mathbf{p})$. For the two cases of Fermi and Bose systems this expression transforms into the Fermi-and Bose-Bessel partition functions
\begin{eqnarray}
\mathcal{Z}_{FB}(\mu,V,T)&=&\prod_\mathbf{p}\Big[1+I_\nu(z)\Big] \\
\mathcal{Z}_{BB}(\mu,V,T)&=&\prod_\mathbf{p}{\Big[1-I_\nu(z)\Big]}^{-1}
\end{eqnarray}
Correspondingly, the equations of state of an ideal Fermi-Bessel and Bose-Bessel gas are
\begin{equation}
\frac{PV}{T}\bigg|^{FB\atop BB}_\nu=\sum_\mathbf{p}\Bigg\{
\begin{array}{cc}
 +\log\Big[1+I_\nu(z)\Big] & ~~ \mathrm{Fermi}   \\
 -\log\Big[1-I_\nu(z)\Big] &  ~~\mathrm{Bose}   
\end{array}
\end{equation}
The average Fermi-Bessel and Bose-Bessel occupation numbers of states then become
\begin{equation}
\langle n_\mathbf{p}\rangle^{FB\atop BB}=\frac{I'_\nu(z)}{1\pm I_\nu(z)} 
=\frac{\big[\nu z^{-1}I_\nu(z)\mp  I_{\nu+1}(z)\big]}{1\pm I_\nu(z)}
\end{equation}
It is easily checked that this function behaves correctly for $z\to0$ in both cases. The apparent divergence at small $z$ is compensated by the factor $z^\nu$ in the small argument expansion of $I_\nu(z)$. This leaves sufficient freedom for chosing the index $\nu$ in order to make the distribution positive. Hence, at a first glance the Fermi-Bessel and Bose-Bessel distributions seem reasonably in accord with the physical requirements. 

The most interesting case is the behaviour at zero temperature $T=0$ or $\beta\to\infty$. We check this for the Fermi-Bessel distribution. Let us first assume that $\epsilon_\mathbf{p}<\mu$ with the chemical potential $\mu>0$ as in the ordinary Fermi distribution. Since in this case $z>0$ is positive tending to $\infty$. This makes $I_\nu$ large with the second term in the nominator dominating which yields $\langle n_\mathbf{p}\rangle^{FB}\sim \nu/z^\frac{3}{2}-1$ negative. Hence there is no occupation below $\epsilon_\mathbf{p}=\mu$. This holds also for finite temperatures. Moreover, for $\epsilon_\mathbf{p}>\mu$ one has $z<0$, and from the asymptotic expansion $\langle n_\mathbf{p}\rangle^{FB}=0$. The above distribution does not exist for $T=0$ and makes also little sense at $T\neq 0$. Similarly the case $T=0$ is excluded for the Bose-Bessel distribution.

Let us now try the function $K_\nu(z)$. It is defined by the integral
\begin{equation}
K_\nu(z)=\int_0^\infty \mathrm{d}t\ \mathrm{e}^{-z\cosh t }\cosh{ \nu t}, \qquad |\arg z|\leq\frac{\pi}{2}
\end{equation}
Its asymptotic expansion is $K_\nu(z)\sim \mathrm{e}^{-z}/\sqrt{2\pi z}$. At $z\to0$ it diverges like $z^{-\zeta}$, where $\zeta=\mathrm{mod}(\nu+1)$. To account for this divergence, one may multiply it with $z^{\zeta}$. One also has
\begin{equation}
K_\nu(-z)=\mathrm{e}^{-i\pi\nu}K_\nu(z)-i\pi I_\nu(z),\qquad \nu\neq 1,2,\dots
\end{equation}
We then have
\begin{equation}
\mathcal{Z}_G(\mu,V,T)=\prod_\mathbf{p}\Big\{\sum_n\Big[z^{\zeta}K_\nu(z)\Big]^n\Big\}
\end{equation}
obtaining
\begin{eqnarray}
\mathcal{Z}_{FB}(\mu,V,T)&=&\prod_\mathbf{p}\Big[1+z^{\zeta}K_\nu(z)\Big] \\
\mathcal{Z}_{BB}(\mu,V,T)&=&\prod_\mathbf{p}{\Big[1-z^{\zeta}K_\nu(z)\Big]}^{-1}
\end{eqnarray}
and for the equations of state
\begin{equation}
\frac{PV}{T}\bigg|^{FB\atop BB}_\nu=\sum_\mathbf{p}\Bigg\{
\begin{array}{cc}
 +\log\Big[1+z^{\zeta}K_\nu(z)\Big] & ~~ \mathrm{Fermi}   \\
 -\log\Big[1-z^{\zeta}K_\nu(z)\Big] &  ~~\mathrm{Bose}   
\end{array}
\end{equation}
The average occupation numbers of states becomes in this case for the Fermi-Bessel and Bose-Bessel distributions
\begin{eqnarray}
\langle n_\mathbf{p}\rangle^{FB\atop BB}&=&\frac{(\zeta+1)z^{\zeta}\big[z^{-1}K_\nu(z)\pm K'_\nu(z)\big]}{1\pm z^{\zeta}K_\nu(z)}\nonumber\\
&=&\frac{\zeta z^{\zeta}\big[(\nu+1) z^{-1}K_\nu(z)\mp K_{\nu+1}(z)\big]}{1\pm z^{\zeta}K_\nu(z)}
\end{eqnarray}
Again both for $z>0$ and $z<0$ at $T=0$ there is no occupation of states! Hence, no correlations can exist at zero temperature. The distributions, if at all, do not exist at $T=0$. 

\section{The classical case}
As we have shown, the use of Bessel functions in order to obtain Bessel-Fermi distributions is not successful. What about the classial case? Does a formal Bessel-Boltzmann distribution exist? The case of the Kappa distribution suggests that this would not be categorically excluded independent on whether the distribution found has any real application to physical problems. So, in the following, we check whether a classical limit exists for Bessel distributions.

The classical case requires that the chemical potential is negative and thus $z=-\beta(|\mu|+\epsilon_\mathbf{p})$ is large. This means that we have to inspect the negative large argument limits of the Bessel functions and their derivatives. 

Let us do this for the modified Bessel function of the first kind. We expect that for large argument the Bessel-Fermi distribution should become the classical equivalent of the Boltzmann distribution similar to the transition from ordinary Fermi to the ordinary Boltzmann distribution. In the limit of very large $|z|\gg1$, where we want to check its validity, we must make use of the large negative argument representation of $I_\nu(z)$, which is the asymptotic expansion of $I_\nu(-\chi)$, with $\chi=|z|$. It yields the following expression
\begin{equation}
{\langle n_\mathbf{p}\rangle_{{\chi\to\infty}}} \simeq \frac{-\nu\mathrm{e}^{\chi+i\pi\nu}/\chi\sqrt{\chi}-\mathrm{e}^{\chi+i\pi(\nu+1)}/\sqrt{\chi}}{1+\mathrm{e}^{i\pi\nu+\chi}/\sqrt{\chi}}
\end{equation}
For large argument $\chi$ the second term in the denominator is much larger than one. Hence, the rest of the denominator shortens with the corresponding parts in the numerator. Moreover, the second term in the numerator changes sign. Thus, in this large argument limit, the result is $\langle n_\mathbf{p}\rangle_{\chi\to\infty} \simeq 1-\nu/\chi$. The classical limit makes obviously sense. Since the last calculation is just its extrem asymptotic value, we obtain
\begin{equation}
\langle n_\mathbf{p}\rangle_\mathit{class}^\mathit{Bessel} = \frac{\nu}{1+\chi}\Bigg[1-\frac{I_{\nu+1}(\chi)}{\nu I_\nu(\chi)}\Bigg], \qquad \nu\neq 0,1,2\dots
\end{equation}
where the argument is $\chi=\beta(|\mu|+\epsilon_\mathbf{p})$. This is a classical Boltzmann-Bessel distribution which corresponds to the Boltzmann distribution. Since $I_\nu(\chi)>I_{\nu+1}(\chi)$, the Boltzmann-Bessel distribution is finite for all $\chi\geq0$. Actually, in this representation the restriction on $\nu$ reduces to $\nu>0$. 

One realises that the factor in front of the brackets is a simple Lorentzian distribution. One thus may note that the last expression can be interpreted as kind of a Bessel-modified Lorentzian distribution of states. 

This suggests various further generalisations. The relativistic version is obtained by mapping $\chi(\epsilon_\mathbf{p})\longmapsto \chi(\gamma)$ with $\epsilon_\mathbf{p} \longmapsto \gamma(\mathbf{p})$ and appropriate redefinition and normalisation of the coefficient $\beta$ and chemical potential $\mu$. Generalisation to generalised Lorentzians is achieved by the replacement of the Lorentzian denominator $1+\chi \to (1+\chi/\kappa)^{(\kappa+s+1)}$ yielding 
\begin{equation}
\langle n_\mathbf{p}\rangle_{\kappa}^\mathit{Bessel} = \frac{\nu}{(1+\chi/\kappa)^{1+s+\kappa}}\Bigg[1-\frac{I_{\nu+1}(\chi)}{\nu I_\nu(\chi)}\Bigg]
\end{equation}
where $s\in\textsf{R}$ is some real number that has to be fixed by bringing the distribution in accord with thermodynamics. Clearly, this is the Kappa distribution multiplied by an additional factor  containing the Bessel functions which serves just as a correction on the generalised Lorentzian.

One could even go further, interpreting the bracket as the expansion of an exponential. This then yields
\begin{equation}
\langle n_\mathbf{p}\rangle_{\kappa, 1}^\mathit{Bessel} = \frac{\nu}{(1+\chi/\kappa)^{1+s+\kappa}}\exp\Bigg[-\frac{I_{\nu+1}(\chi)}{\nu I_\nu(\chi)}\Bigg]
\end{equation}
Finally, further generalisation can be obtained by absorbing the Bessel functions into the generalised Lorentzian in the usual way obtaining
\begin{equation}
\langle n_\mathbf{p}\rangle_{\kappa, 2}^\mathit{Bessel} = \nu\Bigg\{1+\frac{1}{\kappa}\bigg[\chi+\frac{I_{\nu+1}(\chi)}{\nu I_\nu(\chi)}\bigg]\Bigg\}^{-(1+s+\kappa)}
\end{equation}
This is an ordinary Bessel-modified classical Kappa distribution. All these distributions have still to be brought into accord with the thermodynamic requirements.

It is interesting that the functional form of the partition function as the sum over probabilities of states that is obtained from counting of states allows for completely different classical distributions which we have guessed while it suppresses the two quantum distributions. This suppression of the distribution in the quantum domain is not unreasonable because quantum physics relies solely on stochasticity. 

Whether the classical Bessel-Boltzmann distribution and its further generalisations obtained have any physical meaning or not, is a completely different question. We just played with the possibility of a different kind of statistical mechanical distributions of occupation of states arbitrarily chosing Bessel functions for our experiment. Inferring whether a distribution like this one has physical meaning requires the derivation of  the corresponding entropy and testing the thermodynamic relations. 

\section{Conclusion}
We have used the Gibbs-Boltzmann prescription of the partition function in application to different basis factors which replace the so-called Gibbs-Boltzmann factor, the exponential function in the definition of probability. The latter results from the assumption of complete stochasticity in the processes underlying the interaction of the particles respectively systems involved. Their foundation is Gauss' error distribution transformed into energy respectively momentum space. Any replacement of the Gibbs-Boltzmann factor by another more complicated function thus implies that one uses non-stochastic probabilities which may involve correlations. This has been discussed at other places. We have shown that such a replacement works nicely for the generalised Lorentzian factor used in giving the so-called Kappa-distribution a physical fundament. The Kappa-distribution actually becomes a generalised Lorentzian distribution. Its derivation from the generalised partition function results in a slightly different version than used in its otherwise widely distributed %though not completely justified 
applications. Bringing it into complete accord with thermodynamics fixes the free parameter $r$ contained in this distribution to the value as determined in other places \citep[cf.,][]{livadiotis2009,treumann2014b}.

Generalisation of the partition function to the generalised Lorentzian implies that for large $\kappa$ the Lorentzian factor smoothly becomes the Gibbs-Boltzmann factor. We have tentatively dropped this condition and used, as for another physically motivated example, the modified Bessel functions as replacement of the Gibbs-Boltzmann factor. This yields another completely different distribution, which we called Bessel-distribution. Similar to the Lorentzian distribution the two fundamental distributions, the Fermi-Bessel and Bose-Bessel distributions, have no zero temperature limit. This demonstrates again and rather clearly that only the stochastic Gibbs-Boltzmann factor accounts correctly for the zero temperature quantum behaviour. Any other more complicated and non-stochastic distribution necessarily implies the existence of correlations on the level of counting statistics, thus invalidating the distributions on the zero temperature level where no such correlations are allowed because the dynamics is frozen. 

At finite temperatures both distributions might exist. For one of them we have shown that, in the classical domain, it transforms into a reasonable though complicated Boltzmann-Bessel distribution. Whether it has any physical meaning or not, is unknown, however. We do not attempt to check it here as the demonstration intends nothing more than providing an example. 

The new classical distribution turns out to belong to the family of \emph{modified Lorentzian} distributions, the same family to that the Kappa distribution as well as its equivalent in non-extensive $q$-statistical mechanics also belong. It thus seems that the Gibbsian form of the partition function is fundamental not only to Gibbs-Boltzmann statistics but also to all kinds of classical generalised Lorentzians. It obviously includes some particular class of correlations on the probabilistic microscopic  level of states that gives rise to generalised Lorentzian and nonextensive statistical mechanics. 

It would be very interesting in this respect of stepping down into the Gaussian error analysis trying to infer the effect of correlations. One possibility of doing this would be by reference to Bayesian statistics. Bayesian statistics imposes extra conditions -- hypotheses -- which could be physically motivated. Construction of a different Bayesian-Gauss-Gibbs-Boltzmann factor then should provide a physically motivated version of the partition function to be used by standard methods to infer about the resulting average occupation numbers of physical states.    

\begin{acknowledgements}
The present note was part of work on superdiffusion and information theory performed during two short visits of the International Space Science Institute Bern in 2006/2007. RT acknowledges the hospitality of that institution. Moreover, he is particularly indebted to George Livadiotis for his valuable remarks on the manuscript of the present paper, addressing him to the relevant references concerning the Kappa distribution, its history, mathematical and physical contents, and to its various applications in mathematics and space physics.
\end{acknowledgements} 
% For one-column wide figures user
%\begin{figure}
%  \includegraphics{example.eps}
%\end{figure}
%\begin{figure*}
%  \includegraphics[width=0.75\textwidth]{SCALES-B.pdf}
%\end{figure*}

%\begin{table}
%\begin{tabular}{lll}
%\hline\noalign{\smallskip}
%\end{tabular}
%\end{table}

%\begin{acknowledgements}
%\end{acknowledgements}


\end{document}